\documentclass[10pt,conference]{IEEEtran}
\usepackage[dvipsnames]{xcolor} 
\usepackage{amsmath}
\usepackage{amsfonts}
\IEEEoverridecommandlockouts
\usepackage{cite}
\usepackage{algorithm}
\usepackage[utf8]{inputenc}
\usepackage{algpseudocode}
\usepackage{mathtools}
\usepackage{graphicx}
\usepackage{textcomp}
\usepackage{graphicx}
\usepackage{mathrsfs} 
\usepackage{tikz}
\usepackage{pgfplots}
\pgfplotsset{compat=1.18}
\usepackage{subcaption}
\usepackage{pgfplotstable}

\usepackage{caption}
\def\BibTeX{{\rm B\kern-.05em{\sc i\kern-.025em b}\kern-.08em
    T\kern-.1667em\lower.7ex\hbox{E}\kern-.125emX}}
\begin{document}
\title{Leveraging Power Amplifier Distortion\\ for Physical Layer Security
\thanks{This work has received funding from the European Union’s Horizon 2022 research and innovation program under Grant Agreement No 101120332. (EMPOWER-6G project)}
}
\author{
Reza Ghasemi Alavicheh$^*$, Thomas Feys$^*$, MD Arifur Rahman$^\dagger$, François Rottenberg$^*$\\
$^*$ESAT-DRAMCO, KU Leuven, Ghent, Belgium\\
$^\dagger$Research and Innovation Department, IS-Wireless, Piaseczno, Poland\\
\{reza.ghasemialavicheh, thomas.feys, francois.rottenberg\}@kuleuven.be, a.rahman@is-wireless.com
}

\maketitle
\begin{abstract}
This paper introduces a new approach to physical layer security (PLS) by leveraging power amplifier (PA) nonlinear distortion through distortion-aware precoding. While some conventional PLS techniques inject artificial noise orthogonal to legitimate channels, we demonstrate that inherent PA nonlinearities typically considered undesirable can be exploited to enhance security. The zero 3\textsuperscript{rd} order (Z3RO) precoder applies a negative polarity to several antennas to cancel the PA distortion at the user location, resulting in distortion being transmitted in non-user locations. Redirecting the distortion to non-user locations creates interference for potential eavesdroppers, lowering their signal-to-noise-and-distortion ratio (SNDR). Numerical simulations reveal that the Z3RO precoder achieves up to a $2.5\times$ improvement in secrecy rate compared to conventional maximum ratio transmission (MRT) precoding under a $10\%$ outage probability, SNR of $32$ dB and $-5$ dB input back-off (IBO) where the PAs enter the saturation regime.
\vspace{0.05cm}
\end{abstract}
\section{Introduction}


Security in the emerging era of 6G networks, driven by the rapid growth of mobile communications and connected devices, has become critical across numerous applications, from financial transactions to healthcare systems and critical infrastructure. The international telecommunication union (ITU) framework for international mobile telecommunications (IMT)-2030 explicitly identifies enhanced security and resilience as fundamental design principles for future networks \cite{ITU-R_M.2160-0}. While security has traditionally been implemented at upper network layers through cryptographic primitives, emerging network architectures such as ad hoc networks, sensor networks, and internet of things (IoT) devices often face challenges with conventional security approaches due to key management issues or computational limitations \cite{poor_wireless_2017}. This has driven significant research interest in wireless physical layer security (PLS), which exploits the inherent physical aspects of the radio channel to enable secure communications \cite{ara_physical_2024}.
Multiple-antenna systems can improve PLS by directing signals to legitimate users while reducing signal leakage to eavesdroppers \cite{khisti_secure_2010}. The spatial degrees of freedom provided by multiple antennas enable beamforming techniques that concentrate signal energy in specific directions while creating artificial nulls in others, effectively creating a spatially selective communication channel that benefits legitimate users over eavesdroppers. In massive multiple-input multiple-output (MIMO) systems, this capability is further enhanced as the large number of antennas allows for extremely narrow beams with minimal sidelobes, making it increasingly difficult for eavesdroppers to intercept signals unless they are located very close to the legitimate user's direction. Furthermore, the channel hardening effect in massive MIMO systems reduces channel variations, making it easier to reliably predict the performance of security mechanisms and maintain consistent secrecy rates \cite{wu_survey_2018}.

\subsection{State-of-the-Art}
There are various approaches to enhance PLS, from secure coding techniques like low-density parity-check (LDPC) \cite{klinc_ldpc_2011}, polar\cite{yi-peng_wei_polar_2016}, and lattice codes \cite{oggier_lattice_2016} to cooperative methods like cooperative jamming that can degrade the eavesdropper's channel \cite{fakoorian_solutions_2011} and channel-based approaches, including channel-based secret key generation\cite{maurer_secret_1993},\cite{rottenberg_csi-based_2021}, directional modulation\cite{valliappan_antenna_2013}, and secure transmission using reciprocity \cite{ou_secure_2021}. Multiple-antenna systems can also enhance security using a method called antenna subset modulation. This technique uses only a subset of antennas for transmission, with the pattern changing based on a secret key known only to trusted users, making it hard for eavesdroppers to intercept and decode the signal \cite{valliappan_antenna_2013}. Beamforming techniques and artificial noise (AN) generation techniques have emerged as powerful methods to enhance PLS. The basic principle of AN involves projecting artificial noise in the null space of the legitimate channel to degrade the eavesdropper channel without affecting legitimate receivers \cite{goel_guaranteeing_2008}. Several AN methods have been developed, including selfish null space (SNS) AN precoding, which minimizes interference to local cell users, collaborative null space (CNS) AN precoding, which reduces inter-cell AN leakage, computationally efficient polynomial (POLY) AN precoding, and random AN methods that offer lower complexity but higher leakage to legitimate users \cite{zhu_linear_2016}.

Next to this, PA distortion and methods to cope with this distortion in the precoding stage have been widely studied. For instance, in \cite{aghdam_distortion-aware_2021}, a distortion-aware beamforming (DAB) method relying on Bussgang's theorem \cite{bussgang1952} is proposed that maximizes the sum rate in a mmWave multi-user multiple-input single-output (MU-MISO) system. Although promising, this method has a high computational complexity.
Recently, in \cite{moghadam_energy_2018}, it is shown that PA distortion coherently combines at user locations due to beamforming, even with large arrays, which require distortion management in precoding designs. The Z3RO precoders \cite{rottenberg_z3ro_2022, rottenberg_z3ro_2023} mitigate 3\textsuperscript{rd} order distortion through a closed-form and low-complexity solution that saturates selected antennas with negative gain, canceling aggregate distortion while maintaining array gain. Validation based on measured channels \cite{feys_measurement-based_2022} demonstrated distortion reductions of $6.03$ dB and $3.54$ dB in single-user and two-user scenarios, respectively.
For distributed massive MIMO, specialized power allocation methods \cite{liu_power_2022} boost data rates while handling PA nonlinearities. Also, the graph neural networks (GNN) method in~\cite{feys_toward_2024} works with multiple users and complex PA models (i.e., higher order) while staying simpler than DAB precoders. This GNN approach delivers major performance gains ($8.6-8.8$ bits/channel use gain) with better energy efficiency ($3.24\times$ reduction in power consumption) compared to traditional precoding with reasonable computational complexity. Although conventional AN methods inject orthogonal noise into legitimate channels and recent works explore receiver-side distortion cancellation for security \cite{li2024security,wang2024secrecy}, the Z3RO precoder takes advantage of inherent system nonlinearities through transmitter-side spatial precoding to achieve security benefits. By managing PA distortion through specific antenna saturation, the Z3RO precoders can simultaneously improve legitimate signal quality and create channel-specific distortion patterns that function similarly to artificial noise for potential eavesdroppers.
\subsection{Contributions}
To the best of our knowledge, none of the previous works have exploited PA nonlinear distortion to enhance PLS. Our key contributions are as follows
\begin{itemize}
    \item Exploiting PA nonlinear distortion for PLS enhancement and transforming traditionally undesirable hardware impairments into a security feature through the Z3RO precoding technique.
    \item The Z3RO precoder achieves an improvement in secrecy rate up to $2.5\times$ compared to conventional MRT precoding in the saturation regime at $-14$ dB to $0$ dB input back-off (IBO) power and SNR of $32$ dB.
    \item The proposed method produces an effect similar to artificial noise generation without allocating dedicated power resources while improving PA energy efficiency.
\end{itemize}
The benefits are most pronounced in the saturation regime, where PAs operate more efficiently but generate significant nonlinear effects.

The paper is structured as follows: Section~II introduces the system model, including the PA models and the radiation pattern characterization. Section~III presents the secrecy rate definition and outage probability formulation, establishing the key performance metrics for our analysis. In Section~IV, we present the numerical results comparing the Z3RO and MRT precoders across different IBO levels and angular distributions, demonstrating the security advantages of our proposed approach. Finally, Section~V concludes the paper with a summary of key findings and potential directions for future research.
Notation Conventions: Complex space with dimensions $m$ by $n$ is denoted by $\mathbb{C}^{m \times n}$. Scalar values, vectors, and matrices are denoted by $a$, $\mathbf{a}$, and $\mathbf{A}$. $\mathbf{A}^T$ and $\mathbf{A}^H$ indicate a matrix's transpose and Hermitian transpose. The symbol $\mathbb{E}[\cdot]$ represents the expected value operation. The absolute value is denoted by $|\cdot|$. The complex conjugate of a scalar $a$ is denoted by $a^*$.


\section{System Model}
\begin{figure}[tp]
    \centering
    \includegraphics[width=0.5\textwidth]{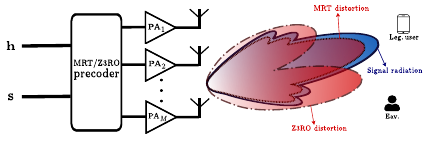}
    \caption{System model showing maximum ratio transmission (MRT) directing signal and distortion to legitimate user, while zero 3\textsuperscript{rd} order (Z3RO) beamforms signal to the user but redirects power amplifier (PA) distortion toward eavesdropper.}
    \label{fig:sysmodel}
\end{figure}

As shown in Fig. \ref{fig:sysmodel}, the system model consists of a base station (BS) equipped with $M$ antennas serving legitimate ($\ell$) and eavesdropper ($e$) users, each with a single antenna assuming a pure line-of-sight (LoS) condition. The channel between BS antenna $m$ and each user $k \in \{\ell, e\}$ is defined as

\begin{equation}
h_{m,k}({\theta}) = \sqrt{\beta_k}e^{-jm\frac{2\pi}{\lambda_c}d\cos(\theta_k)}
\label{eq:LoSchannel}
\end{equation}
where $\beta_k$ represents the path loss, $\lambda_c$ is the carrier wavelength, $d$ is the antenna spacing, and $\theta_k$ is the user angle.
The array response vector is represented as
\begin{equation}
\mathbf{h}_k ({\theta}) = [1, e^{-j\frac{2\pi}{\lambda_c}d\cos({\theta_k})}, \ldots, e^{-j(M-1)\frac{2\pi}{\lambda_c}d\cos({\theta_k})}]^T.
\end{equation}
The function $\phi(\cdot)$ describes the output of a general nonlinear PA and can be written as
\begin{equation}
y = \phi(x(t))
= \phi_A (x(t)) e^{j\angle x(t)+\phi_{\varphi}(x(t))}
\end{equation}
where $\phi_A(\cdot)$ and $\phi_{\varphi}(\cdot)$ denote the amplitude modulation to amplitude modulation (AM/AM) and amplitude modulation to phase modulation (AM/PM) transfer functions, respectively.
Rapp and polynomial models are utilized to model the characteristics of the PA. The modified Rapp model's AM/AM and AM/PM behaviors are described in \cite{feys_toward_2024}
\begin{equation}
\phi_A(x_m) = \frac{A|x_m|}{\left(1 + \left|\frac{x_m}{\sqrt{p_{\text{sat}}}}\right|^{2S}\right)^{\frac{1}{2S}}}
\end{equation}
\begin{equation}
\phi_\phi(x_m) = \frac{|x_m|^q}{1 + \left|\frac{x_m}{B}\right|^q}
\end{equation}
where $p_{\text{sat}}$ is the PA saturation power, $x_m$ is the precoded symbol for a single antenna $m$. $S$ and $q$ are the smoothness factors, $A$ and $B$ are the AM/PM conversion coefficients. IBO is calculated as the ratio of the average input power $p_{\text{in}} = \mathbb{E}[|x_m|^2]$ to the saturation power $p_{\text{sat}}$ as $\mathrm{IBO}$ $= \frac{p_{\text{in}}}{p_{\text{sat}}}$ for each PA.
As a more tractable approximation, a 3\textsuperscript{rd} order polynomial model can be used to describe the PA behavior
\begin{equation}
\phi(x_m) = \beta_1x_m + \beta_3x_m|x_m|^2
\end{equation}
where complex numbers $\beta_1$ and $\beta_3$ are the model's parameters. A least squares regression is carried out to derive these polynomial coefficients at the given IBO by fitting the polynomial model to the Rapp model.

The result of this nonlinear function $\phi(\cdot)$ applied per element of $\mathbf{x}$ can be decomposed using the Bussgang decomposition. The received signal at user $k$ is given by
\begin{equation}
y_k = \mathbf{h}_k^T \boldsymbol{\phi}(\mathbf{x}) + {\nu}_k
\end{equation}
where $\mathbf{h}_k \in \mathbb{C}^{M \times 1}$ is the channel vector based on the LoS channel (\ref{eq:LoSchannel}), $\mathbf{x} \in \mathbb{C}^{M \times 1}$ is the transmitted signal vector, $\nu_k \in \mathbb{C}$ represents complex Gaussian noise with variance $\sigma_{\nu}^2$. The transmitted signal can be decomposed into a linear part and an uncorrelated distortion term, modeled as follows
\begin{equation}
\boldsymbol{\phi}(\mathbf{x}) = \mathbf{G}\mathbf{x} + \mathbf{e}
\end{equation}
where $\mathbf{G} \in \mathbb{C}^{M \times M}$ represents the Bussgang gain matrix, $\mathbf{x} = \mathbf{w}s$ is the precoded signal with $\mathbf{w} \in \mathbb{C}^{M \times 1}$ being the precoding vector, $s \in \mathbb{C}$ the transmitted symbol with \(s \sim \mathcal{CN}(0, 1)\), and $\mathbf{e} \in \mathbb{C}^{M \times 1}$ captures the nonlinear distortion effects. The PA behavior affects the system through the nonlinear uncorrelated distortion term $\mathbf{e}$ in the received signal model.
For the 3\textsuperscript{rd} order polynomial model, the Bussgang gain matrix $\mathbf{G}$ is a function of the precoding vector $\mathbf{w}$ as follows \cite{aghdam_distortion-aware_2021}
\begin{equation}
\mathbf{G}(\mathbf{w}) = \beta_1\mathbf{I}_M + 2\beta_3\text{diag}(\mathbf{w}\mathbf{w}^H).
\end{equation}
The covariance matrix of the distortion $\mathbf{C}_e(\mathbf{w}) \in \mathbb{C}^{M \times M}$ is given by
\begin{equation}
\mathbf{C}_e(\mathbf{w}) = 2|\beta_3|^2(\mathbf{w}\mathbf{w}^H \odot \mathbf{w}^*\mathbf{w}^T \odot \mathbf{w}\mathbf{w}^H).
\end{equation}

The radiation pattern in direction $\tilde{\theta}$ is defined as
\begin{equation}
P(\tilde{\theta}) = \mathbb{E}\left(\left|\sum_{m=0}^{M-1} y_m e^{-jm\frac{2\pi}{\lambda_c}d\cos(\tilde{\theta}
)}\right|^2\right)
\end{equation}
where $y_m$ is the output of the $m$-th PA. Using the Bussgang decomposition \cite{moghadam_energy_2018}, we can characterize the radiation patterns for both signal and distortion components. The signal radiation pattern is $P_{\mathrm{sig}}(\tilde{\theta}) = |{\mathbf{h}(\tilde{\theta})}^T \mathbf{G} \mathbf{w}|^2$ where $\mathbf{h}(\tilde{\theta})$ is the array response vector. The distortion radiation pattern is $P_{\mathrm{dist}}(\tilde{\theta}) = {\mathbf{h}(\tilde{\theta})}^T \mathbf{C}_e \mathbf{h}(\tilde{\theta})^*$. To facilitate meaningful comparison, we define the total transmit power as $P_T = \int_{0}^{\pi} P(\tilde{\theta})d\tilde{\theta}$ and the array directivity as $D(\tilde{\theta}) = \frac{P(\tilde{\theta})}{P_T/\pi}$. The signal component $P_{\mathrm{sig}}(\tilde{\theta})$ and distortion component $P_{\mathrm{dist}}(\tilde{\theta})$ patterns are obtained by evaluating the respective expressions with the appropriate precoding vector $\mathbf{w}$.
\begin{figure}[tp]
    \centering
    \includegraphics[width=0.35\textwidth]{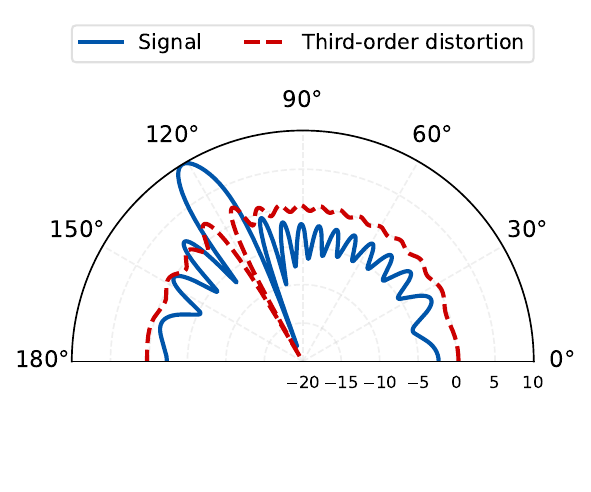} \vspace{-1.0cm}
    \caption{Directivity pattern of signal power and 3\textsuperscript{rd} order distortion for the Z3RO precoder at $\mathrm{IBO}$ = -10 dB. The precoder suppresses distortion at the legitimate user's direction ($120^{\circ}$) while keeping it to other angles to interfere with potential eavesdroppers.}
    \label{fig:z3ro_radiation}
\end{figure}
The maximum ratio transmission (MRT) precoder for legitimate user is formulated as
\begin{equation}
\mathbf{w}_{\text{MRT,$\ell$}} = \frac{\mathbf{h}(\theta_{\text{$\ell$}})^*}{\sqrt{\sum_{m=0}^{M-1} |h_m(\theta_{\text{$\ell$}})|^2}}
\end{equation}
where $h_m(\theta_{\ell})$ is the channel coefficient between antenna $m$ and the legitimate user.
Z3RO, proposed by \cite{rottenberg_z3ro_2023}, is designed to maximize SNR while canceling 3\textsuperscript{rd} order nonlinear distortion at the user location. The Z3RO precoding coefficients are
\begin{equation}
w_{\text{Z3RO},m} = \alpha h^*_m \begin{cases}
-\left(\frac{\sum_{m'=M_s}^{M-1} |h_{m'}|^4}{\sum_{m''=0}^{M_s-1} |h_{m''}|^4}\right)^{1/3}, & \text{if } m \in \mathcal{M} \\
1, & \text{otherwise}.
\end{cases}
\end{equation}
where $\alpha$ is a normalization constant setting the power to one, 
and $\mathcal{M}$ is the set of $M_s$ saturated antennas chosen from the $M$ antennas, with $M/2 > M_s > 0$. The key principle of Z3RO precoding is to saturate a subset of antennas with negative gains to cancel the aggregate 3\textsuperscript{rd} order distortion from all other antennas at the user location. For the special case of the LoS channel, the Z3RO precoder simplifies to
\begin{equation}
w_{\text{Z3RO},m} = \alpha h^*_m \begin{cases}
-\left(\frac{M-M_s}{M_s}\right)^{1/3}, & \text{if } m \in \mathcal{M} \\
1, & \text{otherwise}
\end{cases}
\end{equation}
The key difference between these precoders is that while MRT maximizes the received signal power without considering PA nonlinearities, Z3RO uses a subset of saturated antennas with negative gains to cancel the 3\textsuperscript{rd} order distortion at the legitimate user's location while maintaining a good array gain. This approach is particularly beneficial for large array systems operating near PA saturation.


\section{Secrecy Rate Outage Probability}

The key performance metrics are the signal-to-noise ratio (SNR) and signal-to-noise-and-distortion ratio (SNDR) for each user $k$, expressed as:
\begin{equation} 
\text{SNR}(\theta,\mathbf{w}) = \frac{|\mathbf{h}(\theta)^T\mathbf{G}(\mathbf{w})\mathbf{w}|^2}{\sigma_{\nu}^2}
\end{equation}
\begin{equation} 
\text{SNDR}(\theta,\mathbf{w}) = \frac{|\mathbf{h}(\theta)^T\mathbf{G}(\mathbf{w})\mathbf{w}|^2}{\mathbf{h}(\theta)^T\mathbf{C}_e(\mathbf{w})\mathbf{h}(\theta)^* + \sigma_{\nu}^2}
\end{equation}

where $\mathbf{C}_e(\mathbf{w})$ represents the covariance matrix of the distortion term. The SNR considers only the thermal noise contribution, while SNDR comprehensively captures both thermal noise and nonlinear distortion from the PA on the received signal quality.

\subsection{Secrecy Rate Definition}
Given the SNDR at the legitimate user located at $\theta_{\ell}$ and a potential eavesdropper at angle $\theta$, the instantaneous secrecy rate is defined as\footnote{We make an implicit assumption that the eavesdropper treats distortion as noise and does not attempt to decode or exploit the nonlinear distortion components.}

\begin{equation}
\label{eq:secrecy_rate}
R_s(\theta) = \max\left\{0, \log_2\left(\frac{1 + \text{SNDR}(\theta_{\ell},\mathbf{w})}{1 + \text{SNDR}(\theta,\mathbf{w})}\right)\right\}
\end{equation}
which represents the maximum achievable secure communication rate between the transmitter and the legitimate user in the presence of an eavesdropper at angle $\theta$.

\subsection{Outage Probability Formulation}
The secrecy outage probability for a given target secrecy rate $R_{\text{th}}$ is defined as the probability that the instantaneous secrecy rate goes below this threshold. Considering eavesdroppers uniformly distributed in the angular domain $[0,\pi]$, the secrecy outage probability can be expressed as
\begin{equation}
\label{eq:outage}
P_{\text{out}}(R_{\text{th}}) = \mathbb{P}_{\theta}\{R_s(\theta) < R_{\text{th}}\}
\end{equation}
This probability can be approximated numerically by discretizing the angle space $[0,\pi]$ into $M_{\text{pt}}$ equally spaced points $\{\theta_i\}_{i=1}^{M_{\text{pt}}}$

\begin{equation}
P_{\text{out}}(R_{\text{th}}) \approx \frac{1}{M_{\text{pt}}}\sum_{i=1}^{M_{\text{pt}}} \mathbf{1}_{\{R_s(\theta_i) < R_{\text{th}}\}}
\end{equation}

where $\mathbf{1}_{\{\cdot\}}$ equals 1 when the condition is true and 0 when it's false. This sum calculates what percentage of angular positions have a secrecy rate below the threshold $R{\text{th}}$.

\section{Simulation Results}
\begin{figure}[tp]
    \centering
    \includegraphics[width=0.42\textwidth, height=2.2in]{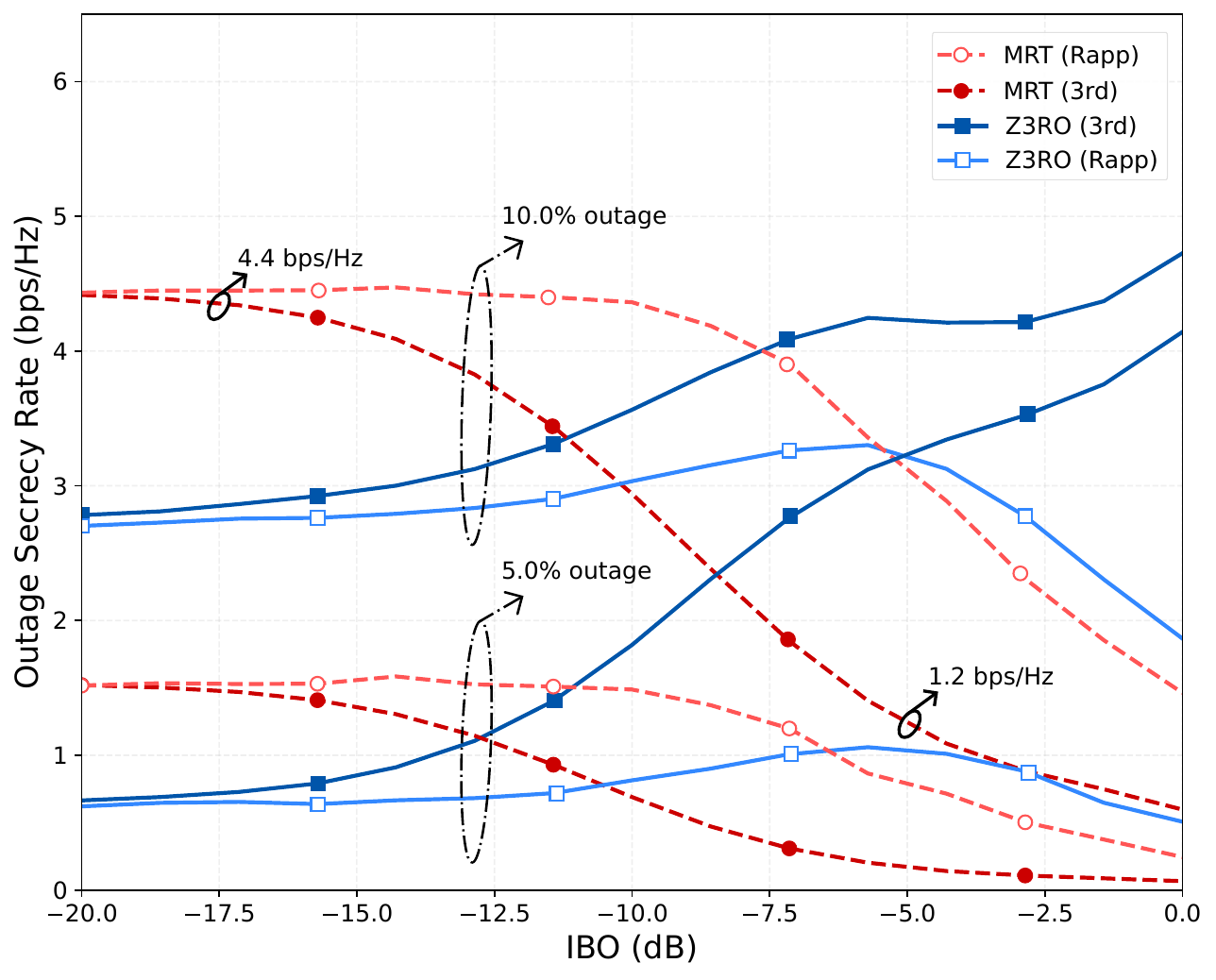}
    \caption{Comparison of secrecy rates between Z3RO and MRT precoders across different $\mathrm{IBO}$ values, considering 3\textsuperscript{rd} order PA models under 5\% and 10\% outage probability.}
     \label{fig:secrecy_rate_vs_ibo_both}
\end{figure}
\begin{figure}[tp]
    \centering
    \includegraphics[width=0.42\textwidth, height=2.2in]{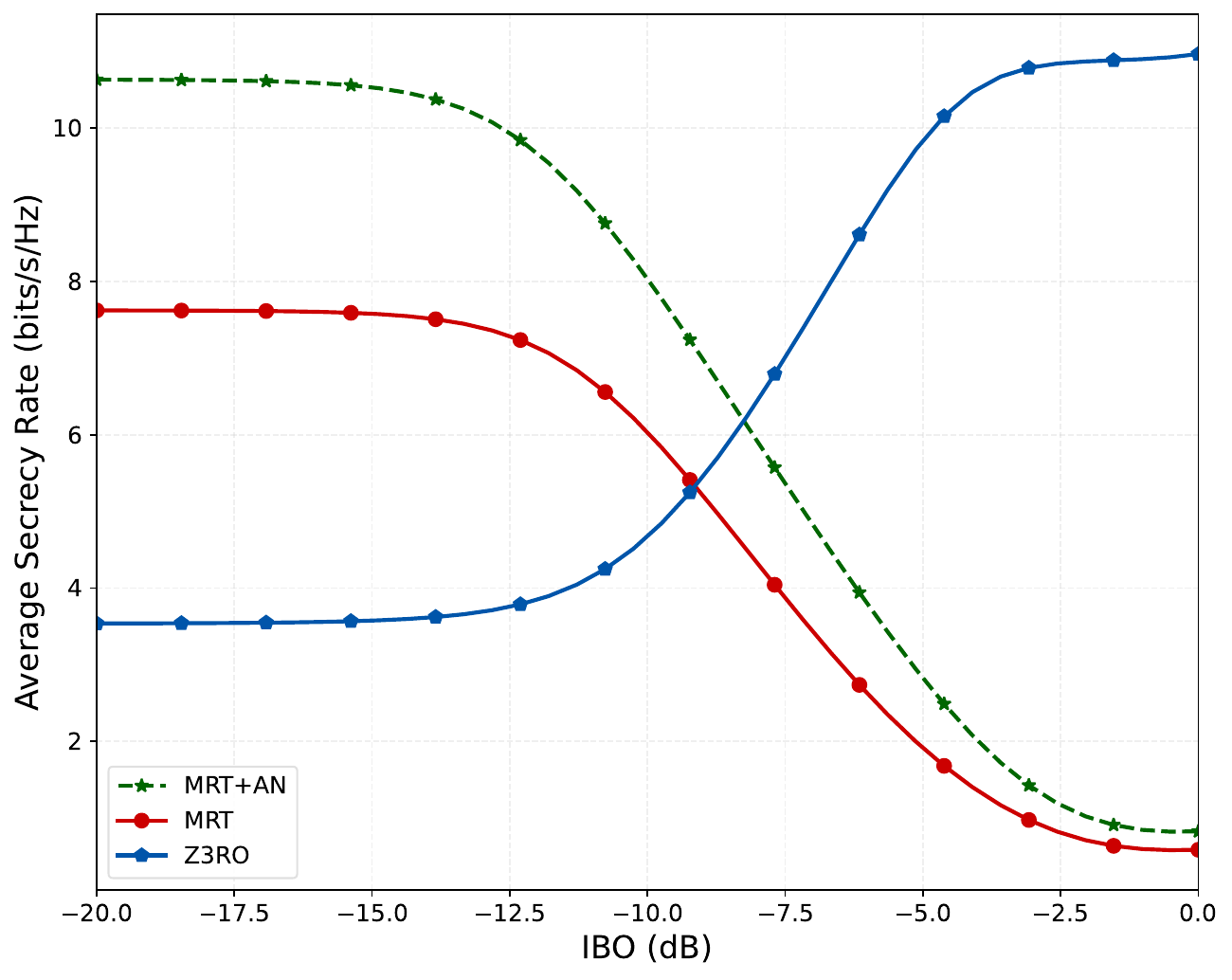} 
    \caption{Average secrecy rate comparison of MRT, Z3RO, and MRT+AN precoding schemes across different input $\mathrm{IBO}$ based on the 3\textsuperscript{rd} Order Polynomial PA model. For the AN scenario, 70\% power allocated to information signal and secrecy rates averaged over all possible eavesdropper locations}
    \label{fig:avg_sec}
\end{figure}
This section evaluates the performance of the proposed Z3RO precoder compared to conventional MRT precoding in terms of secrecy rate. The simulation computes the signal and distortion power distributions across all angular directions by applying the 3\textsuperscript{rd} order PA and Rapp model to MRT and Z3RO precoders. It calculates the Bussgang gain matrix and distortion covariance matrix for each angle and determines the signal and distortion power. These values are used to evaluate the secrecy rates Eq.~(\ref{eq:secrecy_rate}) and outage probabilities Eq.~(\ref{eq:outage}). The simulation parameters are shown in Table~\ref{tab:parameters}.
\begin{table}[t]
\centering
\fontsize{7pt}{8pt}
\caption{Simulation Parameters}
\label{tab:parameters}
\begin{tabular}{|l|l|l|}
\hline
\textbf{Parameter} & \textbf{Description} & \textbf{Value} \\
\hline
$M$ & Number of antennas at base station & 16 \\
\hline
$M_s$ & Number of saturated antennas for Z3RO  & 1 \\
\hline
$\theta_\text{main}$ & Main beam direction (legitimate user) & $120^\circ$ \\
\hline
$\sigma^2_\nu$ & Noise variance & $10^{-2}$ \\
\hline
$\beta$ & Path loss & $1$ \\
\hline
$S$ & Smoothness factor for Rapp AM/AM model & 2 \\
\hline
$q$ & Smoothness factor for Rapp AM/PM model & 4 \\
\hline
$A$ & AM/AM conversion coefficient & -0.315\cite{feys_toward_2024} \\
\hline
$B$ & AM/PM conversion coefficient & 1.137\cite{feys_toward_2024} \\
\hline
$M_\text{pt}$ & Number of angular points& 2000 \\

\hline
\end{tabular}
\end{table}

\begin{figure*}[tp]
    \centering
    \footnotesize  
    \begin{subfigure}[b]{0.32\textwidth}
        \centering
        \includegraphics[width=\textwidth, height=1.5in]{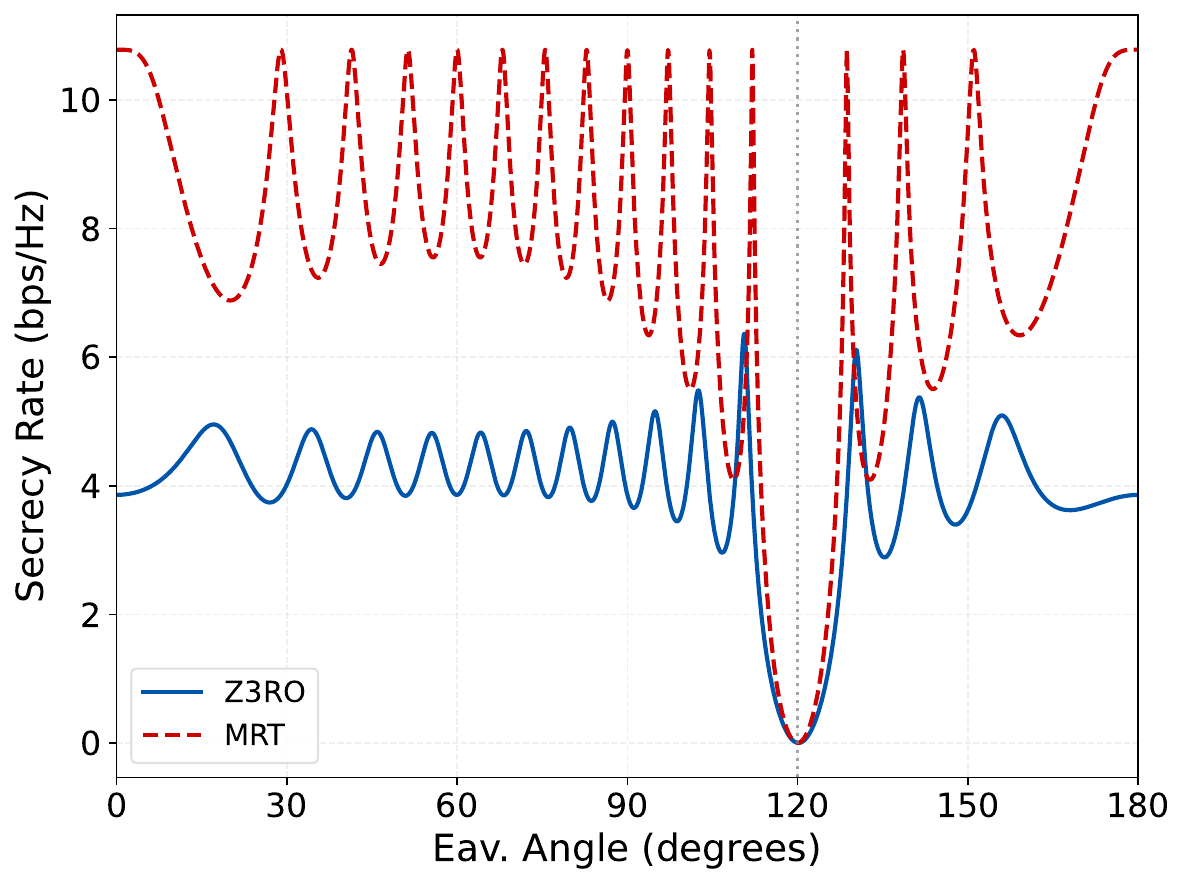}
        \caption{Linear regime (IBO = -20 dB)}
        \label{fig:secrecy_linear}
    \end{subfigure}%
    \hspace{0.01\textwidth}%
    \begin{subfigure}[b]{0.32\textwidth}
        \centering
        \includegraphics[width=\textwidth, height=1.5in]{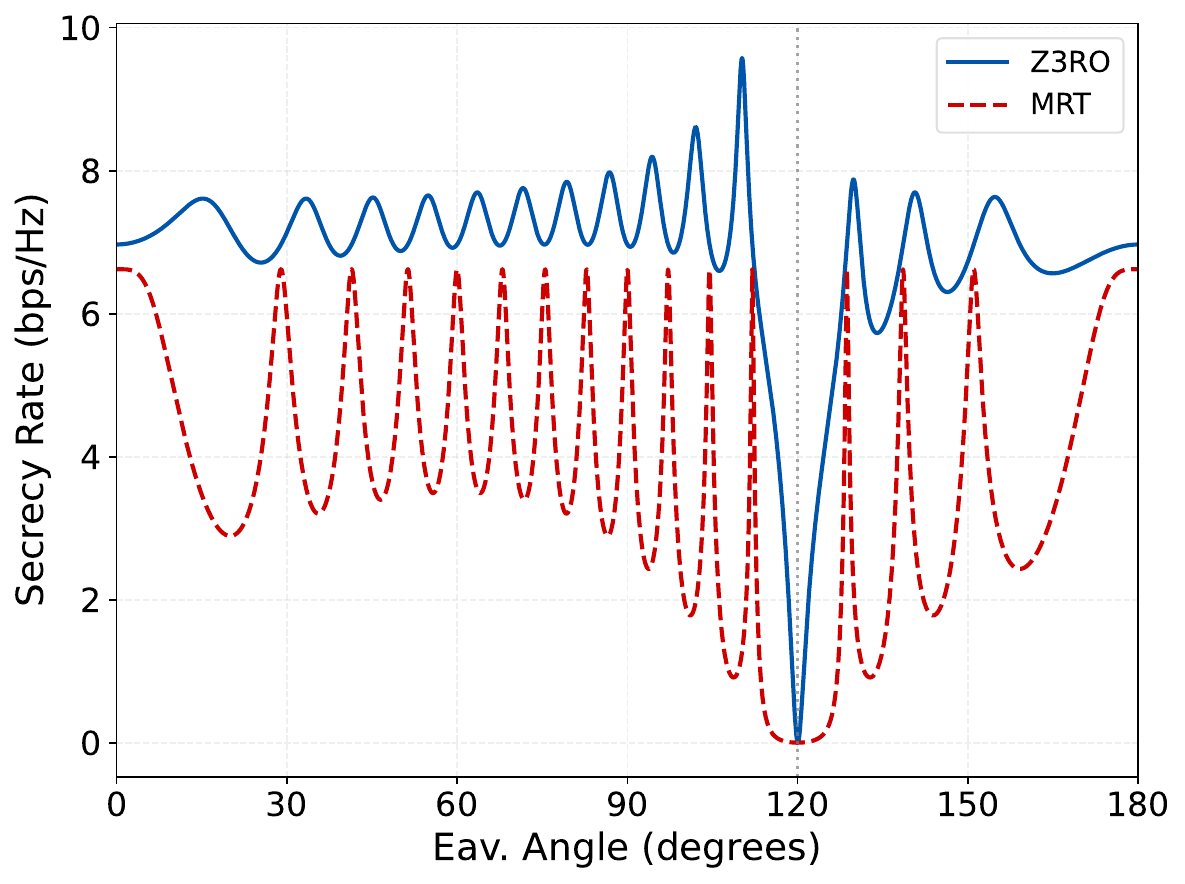}
        \caption{Entering saturation regime (IBO = -10 dB)}
        \label{fig:secrecy_entering}
    \end{subfigure}%
    \hspace{0.01\textwidth}%
    \begin{subfigure}[b]{0.32\textwidth}
        \centering
        \includegraphics[width=\textwidth, height=1.5in]{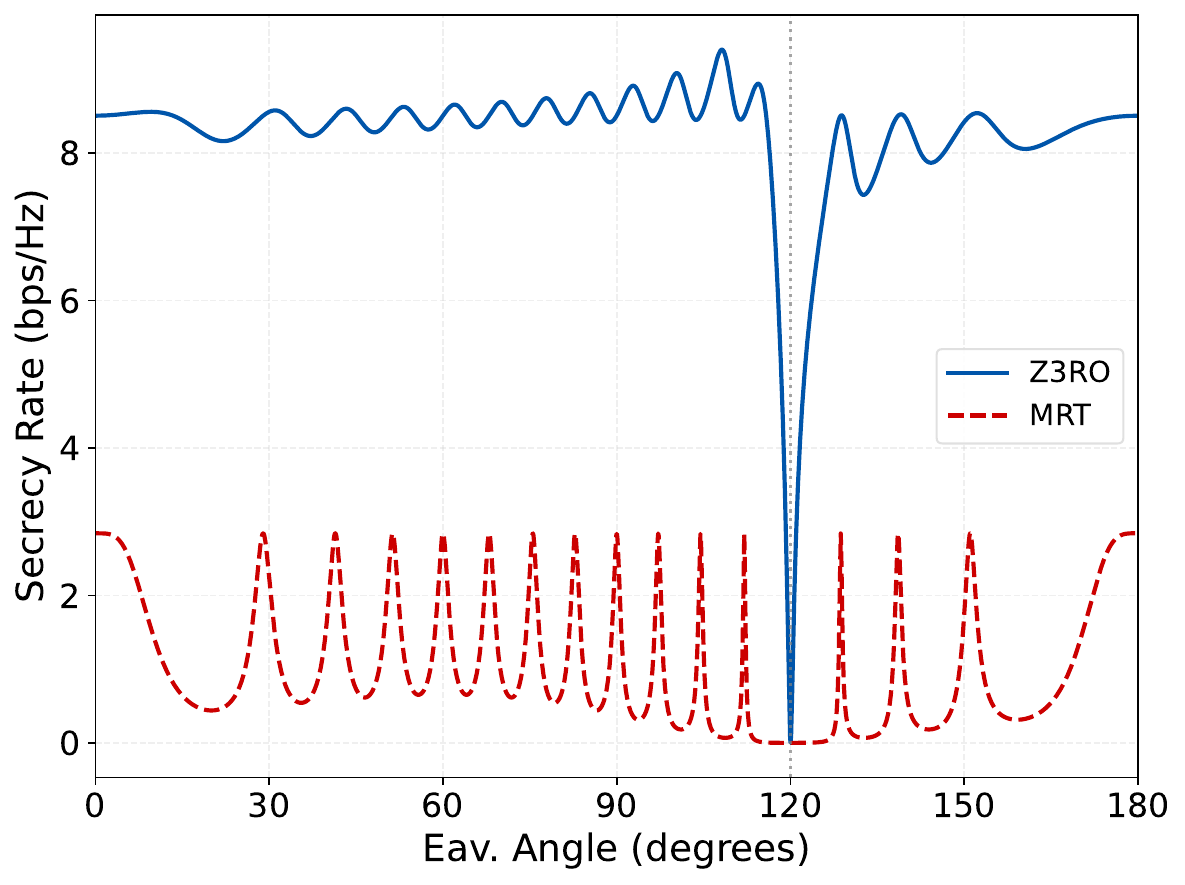}
        \caption{Saturation regime (IBO = 0 dB)}
        \label{fig:secrecy_saturation}
    \end{subfigure}
    \vspace{-0.1cm}
    \caption{Secrecy rate performance across varying eavesdropper angles for different IBO levels.}
    \label{fig:secrecy_rate_ibo_comparison}
\end{figure*}

Fig.~\ref{fig:z3ro_radiation} illustrates the directivity patterns produced by the Z3RO precoder under identical conditions. Based on this figure, the Z3RO precoder not only directs signal power toward the legitimate user but also demonstrates a security-enhancing property as distortion is effectively reduced at the legitimate user's direction ($120^\circ$)
 while being maintained toward other angles. This distortion management directly results from the Z3RO precoder's formulation, which employs negative gain-saturated antennas to interfere destructively with distortion components at the legitimate receiver's location. The distortion pattern at angles away from the legitimate user creates an artificial noise effect that lowers the channel quality for eavesdroppers without harming the legitimate connection. This selective distortion is a key advantage of the Z3RO precoding method.

Fig. \ref{fig:secrecy_rate_vs_ibo_both} demonstrates the secrecy rate performance for both precoders under different PA models and outage constraints (5\% and 10\%). The Z3RO precoder with the 3\textsuperscript{rd} order model shows an improvement in secrecy rate as the system operates deeper into saturation, reaching approximately 4.2 bps/Hz at $\mathrm{IBO}$ = -5 dB under 10\% outage probability. This happens because the precoder can direct nonlinear distortion away from the legitimate user while simultaneously leveraging this distortion as interference against potential eavesdroppers, similar to conventional AN techniques, but without requiring additional power allocation.
Conversely, the MRT precoder shows a significant drop in secrecy rate as amplifiers approach saturation, falling from 4.4 bps/Hz to 1.2 bps/Hz. This represents a decrease of about 3.2 bps/Hz at $\mathrm{IBO}$ = -5 under the 3\textsuperscript{rd} order model. This proves the inability of conventional precoding to exploit nonlinearities for security enhancement.
When using the Rapp model, precoders show different trends compared to the polynomial approximation, as the 3\textsuperscript{rd} order model is only accurate in the early saturation regime but fails deeper in saturation where higher-order terms dominate.
The Z3RO precoder initially maintains high secrecy rates in the saturation regime before experiencing a gradual decline in performance as the system enters deeper saturation.
This decline occurs because Z3RO is specifically designed to cancel 3\textsuperscript{rd} order distortion, while the Rapp model incorporates full PA characteristics, including higher-order nonlinear effects. The 3rd-order model remains accurate when initially entering saturation but becomes less reliable deeper into saturation, where higher-order terms begin to dominate the nonlinear behavior.
Even with this limitation of PA models, Z3RO consistently outperforms MRT across the saturation regime for IBO values larger than approximately -10 dB. The impact of outage probability constraints is evident in the vertical separation between corresponding curves. The more strict 5\% outage requirement results in a 1-3 bps/Hz reduction in achievable secrecy rates compared to the 10\% outage threshold across all precoder and amplifier model combinations. These results demonstrate that the Z3RO precoder effectively transforms the traditionally undesirable nonlinear distortion from PA into a security benefit. This approach delivers dual benefits of security and energy efficiency by effectively utilizing power amplifiers near saturation, where traditional precoding methods typically fail due to increased distortion.
\begin{figure}[tp] 
    \centering
    \includegraphics[width=0.42\textwidth, height=2.2in]{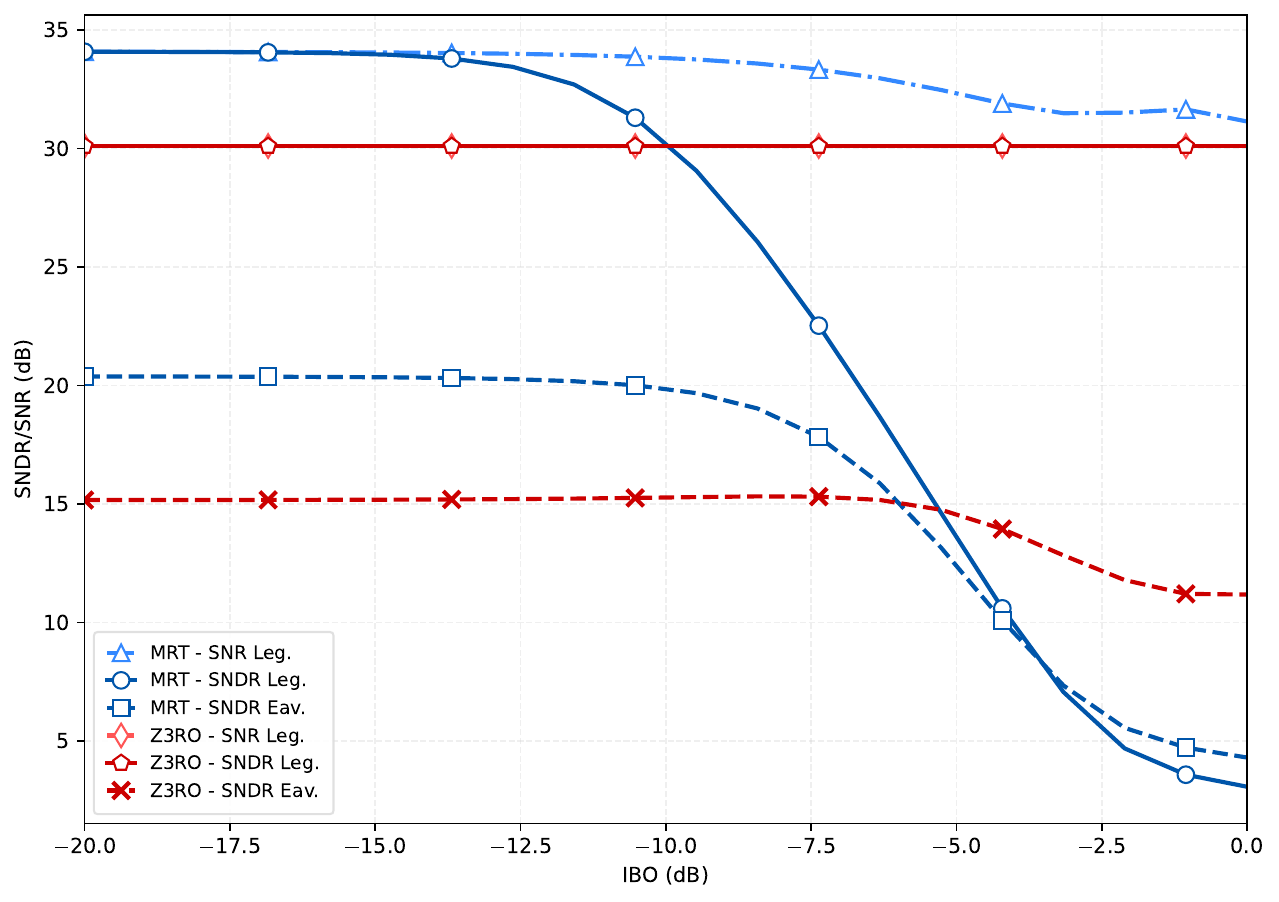}
    \caption{Comparison of Z3RO and MRT precoders showing SNDR and SNR performance for legitimate user and eavesdropper across IBO values based on the 3\textsuperscript{rd} Order Polynomial PA model.}
    \label{fig:sndr_3rd}
\end{figure}
\begin{figure}[tp]
    \centering
    \includegraphics[width=0.42\textwidth, height=2.2in]{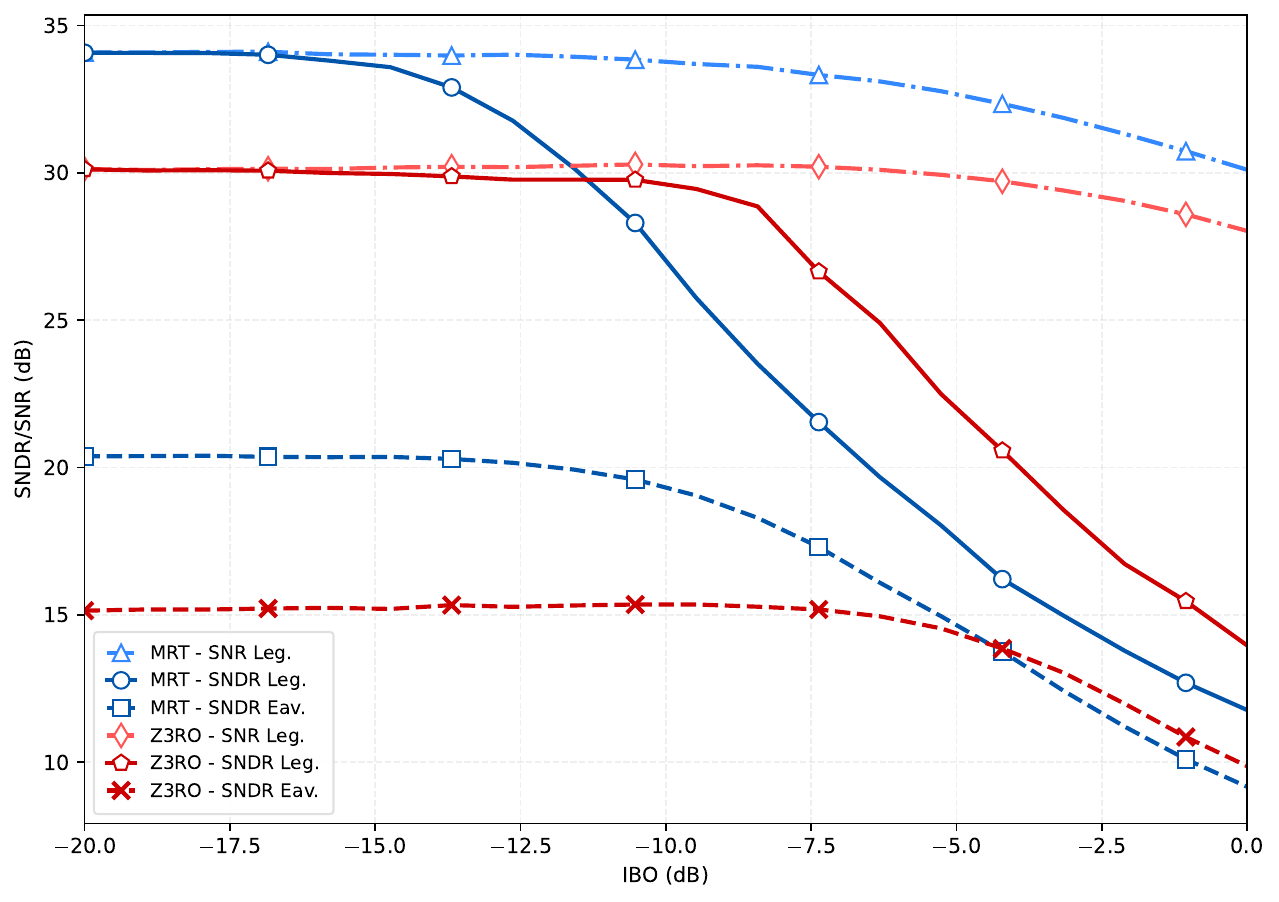}
    \caption{Comparison of Z3RO and MRT precoders showing SNDR and SNR performance for legitimate user and eavesdropper across IBO values based on the Rapp PA model.}
    \label{fig:sndr_rapp}
\end{figure}

Fig.~\ref{fig:avg_sec} shows average secrecy rates as PA nonlinearity increases. At lower $\mathrm{IBO}$ values, AN performs best at 7 bits/s/Hz, using MRT beamforming for information signals with noise transmitted in the legitimate channel's null space. In the saturation regime, Z3RO outperforms MRT and AN, reaching 8.5 bits/s/Hz while MRT falls to 1 bits/s/Hz and AN achieves 2.5 bits/s/Hz at 0 dB IBO. Unlike AN, which requires extra power \cite{goel_guaranteeing_2008}, Z3RO uses inherent PA distortion as interference, enhancing security without additional power.

Fig.~\ref{fig:secrecy_rate_ibo_comparison} illustrates the secrecy rate for different eavesdropper angles under various IBO values. At $\mathrm{IBO}$ = -20 dB, MRT achieves higher secrecy rates in most angular positions, reaching peaks of approximately 10.5 bps/Hz compared to Z3RO's relatively consistent 4-5 bps/Hz. MRT performs better in this regime because the power amplifiers operate in their linear regime. However, as the system operates deeper in saturation ($\mathrm{IBO}$ = -10 dB and 0 dB), Z3RO demonstrates significantly higher secrecy rates (reaching 7-8 bps/Hz at -10 dB and 8-9 bps/Hz at 0 dB) while MRT's performance declines substantially (dropping to 2-3 bps/Hz at 0 dB). This behavior confirms that Z3RO effectively exploits nonlinear PA distortion as a security mechanism by eliminating distortion at the legitimate user's direction ($120^{\circ}$) while directing it toward potential eavesdroppers at other positions.

Figs. \ref{fig:sndr_3rd} and \ref{fig:sndr_rapp} illustrate the comparison of SNDR and SNR performance between Z3RO and MRT precoders for the legitimate user and the potential eavesdropper at various IBO values under both the 3\textsuperscript{rd} order polynomial PA model (Fig. \ref{fig:sndr_3rd}) and Rapp PA model (Fig. \ref{fig:sndr_rapp}), the Z3RO precoder demonstrates superior distortion management. As the system enters saturation (IBO decreasing towards 0 dB), Z3RO maintains significantly higher SNDR for the legitimate user than MRT, which exhibits rapid performance degradation. Simultaneously, Z3RO creates lower SNDR for the eavesdropper, effectively leveraging nonlinear distortion as interference against the unauthorized receiver. These two benefits, which are keeping good signal quality for authorized users while making it worse for the eavesdropper, show how Z3RO turns unwanted PA distortion into a security feature, especially when amplifiers operate more efficiently near their power limits.
    
\section{Conclusion}
In this paper, we explored a new approach to PLS by leveraging undesirable PA nonlinear distortion into a security advantage through the Z3RO precoding technique. Through numerical simulation analysis, we demonstrated several key findings. First, the Z3RO precoder achieves significant secrecy rate improvements (up to $2.5\times$) over conventional MRT precoding in the saturation regime, where PAs operate most energy-efficient. Second, the spatial modification of distortion creates an effect analogous to artificial noise techniques but without requiring dedicated power allocation, simultaneously improving legitimate signal quality while degrading potential eavesdropper channels and improving energy efficiency. This paper establishes a framework for energy-efficient 6G networks by beneficially exploiting hardware nonlinearities rather than trying to avoid them. For future work, extensions to more complex scenarios could be considered. This includes developing precoders using GNN to address higher-order distortion terms and efficiently manage multiple users. Moreover, optimizing the secrecy rate directly rather than relying on Z3RO's distortion cancellation objective can significantly improve PLS performance.

\section*{VI. References}
\renewcommand{\refname}{} 
\bibliographystyle{IEEEtran}
\bibliography{references}
\end{document}